\documentclass[twocolumn,showpacs,preprintnumbers,amsmath,amssymb,aps,prl,floatfix,reprint]{revtex4-1}

\usepackage{graphicx}
\usepackage{dcolumn}
\usepackage{bm}
\usepackage{color}
\usepackage[dvipdfm,colorlinks=false,final,unicode,
	pdftitle={Measurement of the Beta Asymmetry Parameter $A$ in Neutron Beta Decay},
	pdfauthor={D. Mund, B. Maerkisch, M. Deissenroth, J. Krempel, M. Schumann, H. Abele, A. Petoukhov, T. Soldner}]{hyperref}
\usepackage{units}

\newcommand{\perkeo}{\textsc{Perkeo~II}}

\newcommand{\heidelberg}{Physikalisches Institut, Universit{\"a}t Heidelberg, Philosophenweg~12, 69120 Heidelberg, Germany}
\newcommand{\ill}{Institut Laue-Langevin, BP~156, 6,~rue Jules Horowitz, 38042 Grenoble Cedex~9, France}
\newcommand{\ati}{Atominstitut, Technische Universit{\"a}t Wien, Stadionallee~2, 1020 Wien, Austria}

\begin{document}

\title{Determination of the Weak Axial Vector Coupling $\lambda$ =$g_\mathrm{A}/g_\mathrm{V}$ from a Measurement of the $\beta$-Asymmetry Parameter $A$ in Neutron Beta Decay}

\author{D.\,Mund}
\thanks{New family name: \emph{Schirra}.}
\affiliation{\heidelberg}

\author{B.\,M{\"a}rkisch}
\email{maerkisch@physi.uni-heidelberg.de}
\affiliation{\heidelberg}

\author{M.\,Deissenroth}
\affiliation{\heidelberg}

\author{J.\,Krempel}
\altaffiliation[Now at ]{ETH Z{\"u}rich, Institut f{\"u}r Teilchenphysik, Schafmattstr.~20, 8093~Z{\"u}rich, Switzerland}
\affiliation{\heidelberg}

\author{M.\,Schumann}
\altaffiliation[Now at ]{Universit{\"a}t Z{\"u}rich, Winterthurerstr.~190, 8057 Z{\"u}rich, Switzerland}
\affiliation{\heidelberg}

\author{H.\,Abele}
\altaffiliation[Now at ]{\ati}
\email{abele@ati.at}
\affiliation{\heidelberg}

\author{A.\,Petoukhov}
\affiliation{\ill}

\author{T.\,Soldner}
\email{soldner@ill.fr}
\affiliation{\ill}

\date{\today}


\begin{abstract}
We report on a new measurement of the neutron $\beta$-asymmetry parameter $A$ with the instrument \perkeo. Main advancements are the high neutron polarization of $P = 99.7(1)\%$ from a novel arrangement of super mirror polarizers and reduced background from improvements in beam line and shielding. Leading corrections were thus reduced by a factor of 4, pushing them below the level of statistical error and resulting in a significant reduction of systematic uncertainty compared to our previous experiments. From the result $A_0 = -0.11996(58)$, we derive the ratio of the axial-vector to the vector coupling constant $\lambda = g_\mathrm{A}/g_\mathrm{V} = -1.2767(16)$.
\end{abstract}

\maketitle

The Standard Model of weak $V - A$ interactions describes the $\beta^-$ decay of the free neutron $n \rightarrow p + e + \overline{\nu}_\mathrm{e}$ implementing the following parameters: The vector coupling constant $g_\mathrm{V}$ is defined by the product $G_\mathrm{F}V_{\mathrm{ud}}$ of the Fermi constant $G_\mathrm{F}$ = $g_\mathrm{w}^2/M^2_\mathrm{W}$, where $g_\mathrm{w}$ is the electroweak coupling constant and $M_\mathrm{W}$ is the $W$-boson mass, and the matrix element $V_{\mathrm{ud}}$ of the quark mixing Cabbibo-Kobayashi-Maskawa (CKM) matrix. The axial current is renormalized by the strong interaction at low energy. $\lambda$ = $g_\mathrm{A}/g_\mathrm{V}$ is defined as the ratio of the axial vector and vector coupling constants. $\lambda$ is real, if the weak interaction is invariant under time reversal. Searches for time reversal violation can be found in~\cite{Soldner04,Mumm11}.

$\lambda$, $V_{\mathrm{ud}}$, and neutron's lifetime $\tau$ are interconnected by the following equation,
\begin{equation}
\tau^{-1}=C |V_\mathrm{ud}|^2(1+3\lambda^2) f^\mathrm{R}(1+\Delta_\mathrm{R}),
\label{eq:lifetime}
\end{equation}
where $C=G_\mathrm{F}^2 m_\mathrm{e}^5/(2 \pi^3)=1.1613 \times 10^{-4} \unit{s}^{-1}$ in $\hbar=c=1$ units. $f^\mathrm{R}$ is the phase space factor \cite{Wilkinson82,Konrad11} (including the model independent radiative correction) adjusted for the current value of the neutron-proton transition energy. $\Delta_\mathrm{R}$ \cite{Marciano06} is the model dependent radiative correction to the neutron decay rate. Thus $\lambda$ serves as input for a determination of either the CKM matrix-element $V_{\mathrm{ud}}$ or the lifetime $\tau$. The Standard Model requests that the CKM matrix is unitary, a condition which is experimentally tested at the $10^{-4}$ level for the first row~\cite{Hardy09}, and unitary tests are sensitive tools for searches for physics beyond the Standard Model. Previous determinations of $V_{\mathrm{ud}}$ and $V_\mathrm{us}$ raised questions about the unitarity of the CKM-matrix as discussed in~\cite{Towner95,Abele02,Abele03}. Refs.~\cite {Abele08,Severijns11,Dubbers11} list several other motivations for a determination of $\lambda$ and searches for new symmetry concepts in neutron beta decay.
In principle, the ratio $\lambda$ can be determined from QCD lattice gauge theory calculations, but the results of the best calculations vary by up to $30\%$ \cite{Abele08}. The most precise experimental determination is from the $\beta$-asymmetry in neutron decay but
  previous experimental results are not consistent within their
  uncertainties \cite{pdg2011}.

In neutron decay, the probability that an electron is emitted with angle $\vartheta$ with respect to the neutron spin polarization $P = \langle\sigma_z\rangle$ is \cite{Jackson57}
\begin{equation}
W(\vartheta) = 1 +\frac{v}{c}PA\cos(\vartheta),\label{eq:angulardist}
\end{equation}
where $v$ is the electron velocity. ${A}$ is the parity violating $\beta$-asymmetry parameter which depends on $\lambda$. Accounting for order $1\%$ corrections for weak magnetism $A_\mathrm{{\mu}m}$, $g_\mathrm{V}-g_\mathrm{A}$ interference, and nucleon recoil, ${A}$ in Eq.~\eqref{eq:angulardist} reads \cite{Wilkinson82}
\begin{equation}
A = A_0(1+A_\mathrm{{\mu}m}(A_1W_0+A_2W+A_3/W)),
\end{equation}
with  total electron energy $W = E_\mathrm{e} /m_\mathrm{e}c^2+1$ (endpoint $W_0$). The coefficients $A_\mathrm{{\mu}m}$, $A_1$, $A_2$, $A_3$ are from \cite{Wilkinson82} taking a different $\lambda$ convention into consideration. $A_0$ is a function of $\lambda$,
\begin{equation}
A_0=-2\frac{\lambda(\lambda+1)}{1+3\lambda^2},
\end{equation}
where we have assumed that $\lambda$ is real.  In addition, a further small radiative correction \cite{Glueck92} of order $0.1\%$ must be applied.

In this letter, we present a new value for $\lambda$ derived from a measurement of the $\beta$-asymmetry $A$ with the instrument \perkeo\ with strongly reduced systematic corrections and uncertainty. It was installed at the PF1B cold neutron beam position of the Institut Laue-Langevin (ILL) using a highly polarized cold neutron beam. Other correlation coefficients -- the antineutrino-asymmetry parameter $B$~\cite{Schumann07} and the proton-asymmetry parameter $C$~\cite{Schumann08a} -- have been measured at this beam with the same instrument. Neutrons moderated by a cold source were guided via a neutron guide \cite{Haese02,Abele06} to the experiment and were then polarized using two super mirror (SM) coated bender polarizers in crossed (X-SM) geometry \cite{Kreuz05}. An adiabatic fast passage (AFP) flipper allowed to invert the neutron spin direction. After a series of baffles for beam shaping, the transversally polarized neutron beam traversed the \perkeo\ spectrometer and was absorbed in a beam dump. Two beam-line shutters, directly in front and behind the baffles, served to gain information on background~\cite{Abele97,Abele02}.
The main component of the \perkeo\ spectrometer is a split-pair superconducting 1\,T magnet providing $2\times2\pi$ electron guidance from the full fiducial volume to either one of two plastic scintillator detectors with size $440\,\unit{mm} \times 160\,\unit{mm}$ (see the lower sketch in Fig.~1 of~\cite{Reich00}). Details on the spectrometer and electron backscatter suppression can be found in~\cite{Abele97,Abele02}.

From the measured electron spectra $N^\uparrow_i(E_\mathrm{e})$ and $N^\downarrow_i(E_\mathrm{e})$ in the two detectors ($i=1,2$) for neutron spin up and down, respectively, we define the experimental asymmetry as a function of electron kinetic energy $E_\mathrm{e}$ as
\begin{equation}
A_{\mathrm{exp},i}(E_\mathrm{e})=\frac{N^\uparrow_i(E_\mathrm{e}) -
N^\downarrow_i(E_\mathrm{e})}{N^\uparrow_i(E_\mathrm{e}) + N^\downarrow_i(E_\mathrm{e})}.
\end{equation}
This experimental asymmetry is directly related to the asymmetry parameter $A$, as follows from Eq.~\eqref{eq:angulardist} and $\langle\cos(\vartheta)\rangle = 1/2$:
\begin{equation}\label{eq:asyfit}
A_{\mathrm{exp}}(E_\mathrm{e}) = \frac{1}{2}\frac{v}{c} A P f,
\end{equation}

with neutron polarization $\textit{P}$ and spin flip efficiency $\textit{f}$.

The main experimental errors of this measurement are due to \emph{statistics}, \emph{detector response}, \emph{neutron spin polarization}, and \emph{background subtraction}, see Tab. \ref{tab:errors}. The fourfold intensity of the PF1B~beam compared to the previous PF1~beam is used to enhance both statistics and systematics. The detected neutron decay rate within the fiducial volume was $375\,\unit{s^{-1}}$.

\begin{figure}[htb]
\centering \includegraphics[width=\linewidth, clip=true]{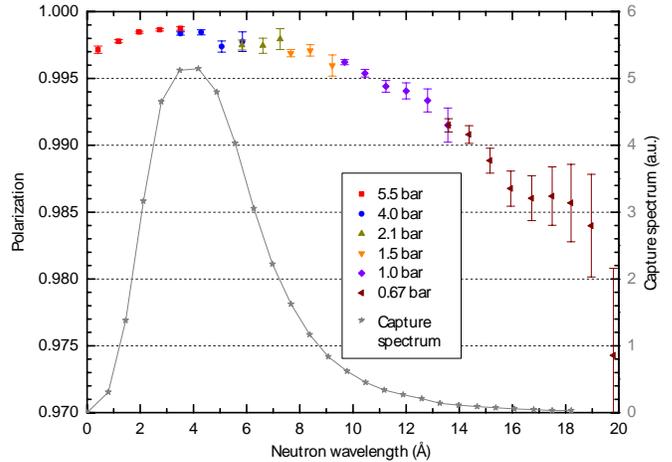}
\caption{Neutron polarization in the center of the \perkeo\ beam. In order to cover the full spectrum with opaque \textsuperscript{3}He cells, 6 cells were used, with lengths of 25\,cm or 10\,cm and different pressures. In the legend, the effective pressures for a 10\,cm cell are given. (Color online) \label{fig:pol}}
\end{figure}

\emph{Polarization:} The X-SM geometry efficiently suppresses garland reflections, resulting in a nearly wave\-length- and angle-independent beam polarization. This dramatically reduces systematic uncertainties for determining the average beam polarization. Polarization measurements were performed employing time-of-flight behind a chopper to gain wavelength resolution. A second AFP flipper and two Sch{\"a}rpf polarizers \cite{Schaerpf89,Schaerpf89a} in X-SM geometry as analyzers were used to measure the spin flip efficiency and for a rough determination of the beam polarization. Measurements in front and behind the \perkeo\ spectrometer yielded consistent results. The absolute polarization was determined using a series of opaque \textsuperscript{3}He spin filter cells of different pressures and lengths, covering the wavelength range from $2\,\unit{\AA}$ to $20\,\unit{\AA}$, see Fig.~\ref{fig:pol}. Cells with both orientations of the \textsuperscript{3}He spin were used to increase sensitivity \cite{Zimmer99a}. Wavelength averages were calculated taking into account the decay probability which is proportional to the measured capture spectrum. The spatial dependence was verified by measurements at five different positions across the neutron beam and found to be negligible. The resulting averages were $P=99.7(1)\%$ and $f=100.0(1)\%$. Note that opaque \textsuperscript{3}He spin filters have an intrinsic accuracy of better than $10^{-4}$ for polarization analysis \cite{Soldner11}.

\begin{figure}[htb]
\centering \includegraphics[clip=true]{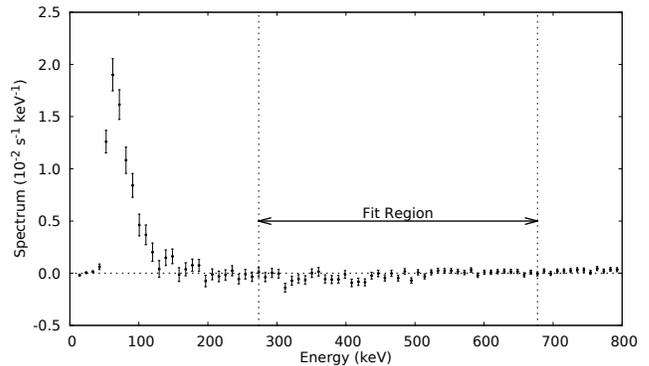}
\caption{Difference of spectra in detector~1 for the second and the first beam line shutter closed, a measure for the background produced by the collimation system. The main part of the background is low energetic. In detector~2, the background looks similar.
\label{fig:bg}}
\end{figure}

\emph{Background:} The magnetic field of \perkeo\ collects all electrons from the decay volume ($100 \times 80 \times 270\,\unit{mm^3}$) and thus assures a high signal-to-background ratio, together with thin plastic scintillators (5\,mm thickness). Environmental background was reduced by lead and iron shielding and measured with the first beam-line shutter closed. It was not constant in time due to shutter operations of neighboring instruments. Changes were monitored by NaI scintillators placed outside the \perkeo\ shielding, and shutter operations were registered. For the analysis, only data sets with constant environmental background were used. This procedure reduces the data set to $5.9 \times 10^7$ neutron decay events and increases the relative statistical error to $3.8 \times 10^{-3}$ (compared to $2.6 \times 10^{-3}$ in the preliminary analysis of~\cite{Mund06,Abele08}). The trigger rate of $500\,\unit{s^{-1}}$ comprises about $120\,\unit{s^{-1}}$ from environmental background. The signal-to-background ratio in the fit region was better than $8:1$.

Beam-related background is more difficult to address. In the \perkeo\ spectrometer, the $\beta$-detectors are far off the beam at a transverse distance of $960\,\unit{mm}$. The beam line was optimized to place the last beam-defining baffle further away from the spectrometer than it was in our previous measurement \cite{Abele02}. The beam stop was positioned $4\,\unit{m}$ downstream of the decay volume. Baffles and beam stop were made from enriched \textsuperscript{6}LiF ceramics. The baffles' lead supports shielded capture gammas (about $10^{-4}$ per capture). Supports and beam line were protected against scattered neutrons by \textsuperscript{6}LiF rubber or boron glass. Halo baffles (not touching the beam) absorbed scattered neutrons close to the beam. Lead shielding was placed around the spectrometer to assure that gamma rays are scattered at least twice before they can reach a detector. In \textsuperscript{6}LiF, about $10^{-4}$ fast neutrons are produced per capture from (t,\,n) reactions \cite{Lone80}. These neutrons were shielded by borated polyethylene (or,  inside the beam stop vacuum, Plexiglas surrounded by borated glass) and secondary gammas by lead. The beam-related background was estimated from measurements with the two shutters using an extrapolation procedure described in \cite{Reich00} and confirmed by additional tests with external background sources. Compared to the previous measurement \cite{Abele02} of 1/200, it was reduced to 1/1700 of the electron rate in the fit region, which corresponds to $0.11(2)\,\unit{s^{-1}}$, see Fig.~\ref{fig:bg}, resulting in a correction of $1(1) \times 10^{-3}$. The assumed relative uncertainty of $100\%$ is a very conservative estimate for this background extrapolation method.

\begin{figure}[t]
\begin{center}
\includegraphics[clip=true]{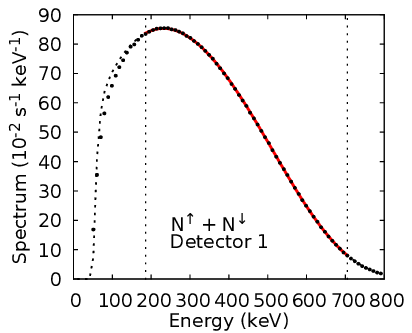}
\includegraphics[clip=true]{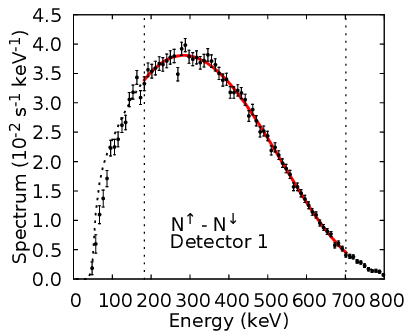}
\includegraphics[clip=true]{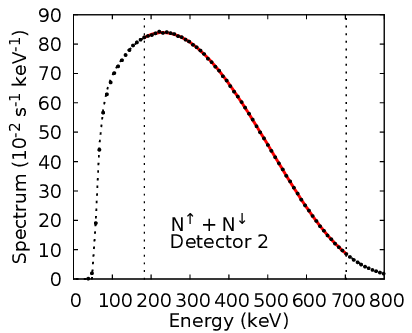}
\includegraphics[clip=true]{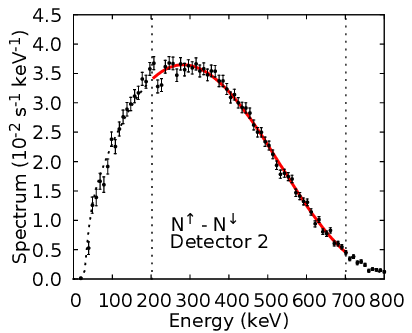}
\end{center}
\caption{Neutron $\beta^-$-spectra after adding or subtracting opposite neutron spin orientations $N^\uparrow$ and $N^\downarrow$ for detector~1 and detector~2, respectively. The solid curve shows a fit to the spectra in the range indicated by the vertical lines. The spectra $(N^\uparrow - N^\downarrow)$ are per se free of background.
\label{fig:spectra}}
\end{figure}

\begin{figure}[t]
\begin{center}
\includegraphics[clip=true]{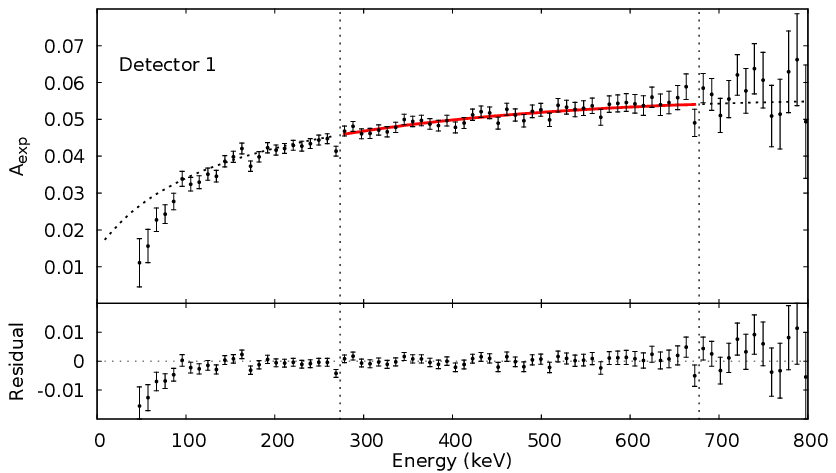}
\includegraphics[clip=true]{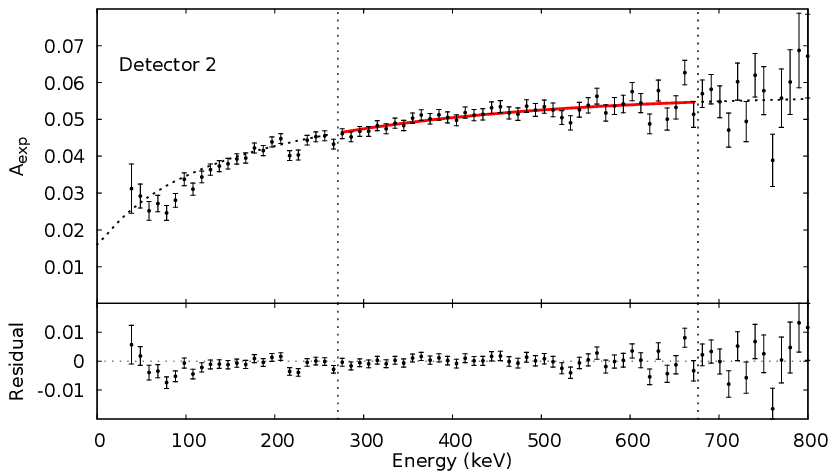}
\end{center}
\caption{Fit and residuals to the experimental asymmetry $A_\mathrm{exp}$ for detector~1 and detector~2. The solid curve shows the fit interval, whereas the dotted curve shows an extrapolation to higher and lower energies.\label{fig:fit}}
\end{figure}

\emph{Detector response:} The plastic scintillators were read out by four photomultipliers per detector. Signals were integrated by charge-to-digital converters over a time interval that includes signals from backscattering. Trigger time differences between the detectors were registered to attribute the event in case of backscattering. The detector response function was determined and the detector stability was checked regularly using four mono-energetic conversion electron sources (\textsuperscript{109}Cd, \textsuperscript{113}Sn, \textsuperscript{207}Bi, and \textsuperscript{137}Cs) on $10\,\unit{\mu g/cm^2}$ carbon backings, which were remotely inserted into the spectrometer. The branching ratios for K, L, M and N conversion electrons and the corresponding Auger electrons have been measured separately with silicon detectors~\cite{Abele93} and were taken into account in the corresponding fit functions. Drift in the detector gain was smaller than $1\%$ and corrected for. The detectors showed a small non-linearity at low energy. The largest systematic uncertainty is caused by the spatial non-uniformity of the detector response. Collected light output for electrons detected in the center of the scintillator was about $5\%$ lower than for electrons detected at the ends. This spatial dependence was mapped using different calibration sources and was found to follow the expected $\cosh$ dependence. The detector gain, which has been used for the fit to the asymmetry parameter~$A$ of Eq.~\eqref{eq:asyfit} and Fig.~\ref{fig:fit}, was obtained by a fit to the spectrum $(N^\uparrow - N^\downarrow)$, see Fig.~\ref{fig:spectra}, which is background free. A fit to $(N^\uparrow + N^\downarrow)$ would yield a different gain resulting in a significant dependence of the asymmetry $A$ on the lower limit of the fit region for detector~2.

\begin{table}[t]
\begin{center}
\begin{ruledtabular}
\begin{tabular}{ldd}
Type & \multicolumn{1}{c}{Correction} & \multicolumn{1}{c}{Uncertainty} \\
     & \multicolumn{1}{c}{($10^{-3}$)} & \multicolumn{1}{c}{($10^{-3}$)} \\
\hline
 Neutron polarization                                 & 3.0 & 1.0 \\
 Spin flip efficiency                                 & 0.0 & 1.0 \\
 Background                                           & 1.0 & 1.0 \\
 Detector response                                    & 0.0 & 2.5 \\
 Electron backscattering                              & 0.25 & 0.04 \\
 Edge effect (1)                                      & (-1.6) & 0.5 \\
 Magnetic mirror effect                               & 0.6 & 0.2 \\
 Dead time (2)                                        & (-1.2) & 0.1 \\[5pt]
 Radiative Correction                                 & 0.9 & 0.5 \\[5pt]
 Statistics                                           & 0.0 & 3.8 \\
\end{tabular}
\end{ruledtabular}
\end{center}
\caption{Summary of corrections and uncertainties to the beta asymmetry $\Delta A_0/A_0$. (1) is included in the fit function, (2) is measured by the data acquisition system and accounted for in the data set.}
\label{tab:errors}
\end{table}

\emph{Backscattering:} By using the second detector as veto detector and carefully analyzing events with energies below the trigger threshold on one detector \cite{Schumann08b}, the fraction of wrongly attributed electrons in the fit region was deduced to be $1.3(3) \times 10^{-4}$ per detector, corresponding to a correction of $0.25(4) \times 10^{-3}$ compared to $2.0(1.7) \times 10^{-3}$ in the previous experiment \cite{Abele02}.

\emph{Edge effect:} The length of the decay volume is defined by electron absorbing aluminum baffles. The absorption depends on the radius of gyration and thus on the energy of the electron. The resulting correction can be reliably calculated and is included in the fit function.

\emph{Mirror effect:} Electrons can be reflected on an increasing magnetic field, leading to detection in the wrong detector. This magnetic mirror effect, due to a small displacement between neutron beam and the maximum of the magnetic field, caused a difference of $1.4$\% between the asymmetries measured in the two detectors. Most of this effect cancels by averaging the two detectors. The remaining correction, due to the spacial extension of the neutron beam, was calculated from the measured magnetic field geometry and neutron beam profile.


The experimental function $A_{\mathrm{exp},i}(E_\mathrm{e})$ and a fit with a single free parameter $\lambda$ are shown in Fig.~\ref{fig:fit} for both detectors. The fit interval was chosen such as to minimize effects due to non-linearity of the detector and unrecognized background.
From the experimental asymmetries we get $|A_\mathrm{0}| = 0.11846(64)_\mathrm{stat}$ for detector~1 and $|A_\mathrm{0}| = 0.12008(64)_\mathrm{stat}$ for detector~2. All subsequent corrections and uncertainties entering the final determination of $A_0$ are listed in Tab.~\ref{tab:errors}. After averaging and correcting for those small and mostly experimental systematic effects, we obtain
\begin{align}
A_0 & = -0.11996(45)_\mathrm{stat}(37)_\mathrm{sys} = -0.11996(58)~ \text{and} \nonumber \\
\lambda & = -1.2767(16).
\end{align}
This value is consistent with our earlier result \cite{Abele02} of $A_0 = -0.1189(7)$. The combined results of the new and our previous \cite{Abele97,Abele02} experiments are
\begin{equation}
A_0 = -0.11951(50)~ \text{and} ~\lambda = -1.2755(13)\label{eq:lambda}.
\end{equation}
In the average Eq.~(\ref{eq:lambda}) we have accounted for correlations of systematic errors in the experiments. Conservatively, errors concerning detector calibration and uniformity, background determination, edge effect and the radiative correction were considered correlated on the level of the smallest error of all three experiments.

Other experiments \cite{Yero97,Liaud97,Bopp86} gave significantly lower values for $|\lambda|$. However, in all these experiments large corrections had to be applied for neutron polarization, magnetic mirror effects, solid angle, or background, which were in the $15\%$ to $30\%$ range. In our present experiment, all individual corrections are below $3 \times 10^{-3}$ and the sum of their absolute values is below $1\%$. We therefore use only the value given in Eq.~\eqref{eq:lambda} for further discussion. The determination of $\lambda = -1.27590_{-0.00445}^{+0.00409}$ by the UCNA collaboration \cite{Liu10} is in agreement with this result, albeit with a larger error.

Assuming the $V-A$ structure of the Standard Model, neutron lifetime can be determined using the $\overline{\mathcal{F}t}$ value from nuclear beta decay
\begin{equation}
\tau_\mathrm{n} = \frac{2}{\ln 2} \frac{\overline{\mathcal{F}t}}{f_R\,(1+3\lambda^2)},
\end{equation}
where the phase space factor $f_R = 1.71385(34)$ \cite{Konrad11} includes radiative corrections. We use our result Eq.~\eqref{eq:lambda} and $\overline{\mathcal{F}t} = 3071.81(83)$ \cite{Hardy09} to derive the neutron lifetime
\begin{equation}
\tau_\mathrm{n} = 879.4(1.6)\,\unit{s}.
\label{eq:ourlifetime}
\end{equation}
This result is in agreement with and nearly as precise as the current world average $\tau_{\mathrm{n}} = 881.9(1.3)\,\unit{s}$ \cite{Dubbers11} that includes a scale factor of $2.5$.

\begin{acknowledgments}
The authors would like to thank Markus Brehm for his contribution during the preparation of the measurement and Anthony Hillaret for his contribution during the beam time, and several services of the Physikalisches Institut, University of Heidelberg, and of the ILL, in particular the Neutron Optics Group.
This work was supported by the Priority Programme SPP~1491 of the Austrian FWF and the German DFG, contracts FWF I529-N20, \mbox{MA 4944/1-1} and \mbox{SO 1058/1-1}, and the German BMBF, contract No.~06HD187.
\end{acknowledgments}

%

\end{document}